\documentclass[article]{aa}
\usepackage{graphicx}
\usepackage{longtable}
\usepackage{lscape}
\usepackage{siunitx}
\usepackage{float}
\usepackage{natbib}
\usepackage[draft]{hyperref}
\usepackage{appendix}
\usepackage{subcaption}
\usepackage{longtable}
\hypersetup{
    colorlinks = true,
    linkcolor = {blue},
    citecolor = {blue},
}

\title{Time evolution of magnetic activity cycles in young suns: The curious case of $\kappa$ Ceti}
\author{S. Boro Saikia \inst{1},~T.~L\"uftinger \inst{2},~C.~P.~Folsom \inst{3,4}, ~A.~Antonova \inst{5},
~E.~Alecian \inst{6}, J.~-F.~Donati \inst{7,8}, 
M.~Guedel \inst{1}, J.~C.~Hall \inst{9},~S.~V.~Jeffers \inst{10},~O.~Kochukhov \inst{11}, 
S.~C.~Marsden \inst{12},~Y.~T.~Metodieva \inst{1},~M.~Mittag \inst{13},~J.~Morin \inst{14}, 
V.~Perdelwitz \inst{13, 15},~P.~Petit \inst{7, 8},~M.~Schmid \inst{1},~A.~A.~Vidotto \inst{16,17}
}
\institute{University of Vienna, Department of Astrophysics,
              T\"urkenschanzstrasse 17, 1180 Vienna, Austria \label{inst1} \and
European Space Agency, European Space Research and Technology Centre, Keplerlaan 1, 2201-AZ Noordwijk, The Netherlands  \label{inst2} \and
Department of Physics and Space Science, Royal Military College of Canada, PO Box 17000 Station Forces, Kingston, ON, Canada K7K 0C6 \label{inst3} \and
Tartu Observatory, University of Tartu, Observatooriumi 1, Tõravere, 61602 Tartumaa, Estonia\label{inst4} \and
Institute of Astronomy and NAO, Bulgarian Academy of Sciences, 72 Tsarigradsko Chaussee Blvd., 1784 Sofia, Bulgaria \label{inst5} \and
Universit\'e Grenoble Alpes, CNRS, IPAG, 38000 Grenoble, France \label{inst6} \and
Universit\'e de Toulouse, UPS-OMP, IRAP, Toulouse, France \label{inst7} \and
CNRS, Institut de Recherche en Astrophysique et Planetologie, 14 avenue Edouard Belin, 31400 Toulouse, France \label{inst8} \and
Lowell Observatory, 1400 West Mars Hill Road, Flagstaff, AZ 86001, USA \label{inst9} \and
Max Planck Institut f\"ur Sonnensystemforschung, Justus von Liebig Weg 3, 37077 G\"ottingen, Germany \label{inst10} \and
Department of Physics and Astronomy, Uppsala University, Box 516, SE-75120 Uppsala, Sweden \label{inst11} \and
University of Southern Queensland,Centre for Astrophysics, Toowoomba, QLD 4350, Australia \label{inst12} \and
Hamburger Sternwarte, Universit\"at Hamburg, Gojenbergsweg 112, 21029 Hamburg, Germany \label{inst13} \and
LUPM-UMR 5299, CNRS $\&$ Universit\'e Montpellier, place Eugène Bataillon, 34095 Montpellier Cedex 05, France \label{inst14} \and
Department of Physics, Ariel University, Ariel 40700, Israel \label{inst15} \and
School of Physics, Trinity College Dublin, University of Dublin, Dublin-2, Ireland \label{inst16} \and
Leiden Observatory, Leiden University, PO Box 9513, 2300 RA Leiden, The Netherlands \label{inst17}
}
\titlerunning{$\kappa$ Ceti}
\authorrunning{S.Boro Saikia et al. }
\abstract{
A detailed investigation of the magnetic properties of 
young Sun-like stars can provide valuable information on our Sun's magnetic
past and its impact on the early Earth.} {We determine the properties of 
the moderately rotating young Sun-like star $\kappa$ Ceti's 
magnetic and activity cycles using 50 years
of chromospheric activity data and six epochs of spectropolarimetric observations.}
{The chromospheric activity was determined by measuring the
flux in the \ion{Ca}{II} H and K lines. A generalised Lomb-Scargle
periodogram and a wavelet decomposition were used on the 
chromospheric activity data to establish the associated periodicities. 
The vector magnetic field of the star was reconstructed 
using the technique of Zeeman Doppler imaging on the spectropolarimetric observations.}
{Our period analysis algorithms detect a 3.1 year chromospheric cycle in addition to the star's
well-known $\sim$~6 year cycle period. Although the two cycle periods have an approximate 1:2 ratio, 
they exhibit an unusual temporal evolution. Additionally, the spectropolarimetric
data analysis shows polarity reversals of the star's large-scale magnetic field,
suggesting a $\sim$10 year magnetic or Hale cycle.}{The unusual evolution of the
star's chromospheric cycles and their lack of a direct correlation with the 
magnetic cycle establishes $\kappa$ Ceti as a curious young Sun. 
Such complex evolution of magnetic activity
could be synonymous with moderately active young Suns, 
which is an evolutionary path that our own Sun
could have taken.

} 
\begin{document}
\maketitle
\section{Introduction}
Young Sun-like stars provide a unique window into our 
solar environment during the first few hundred million years. 
The young Sun
$\kappa$ Ceti is particularly interesting, as its moderate
rotation rate suggests it falls in the rotational evolution track
that our Sun could have taken in the past \citep{lammer20,johnstone21b}.
This star was investigated as
part of the `Sun in Time' project \citep{guedel97,ribas05}
which studied stellar magnetic activity in the 
X-ray and ultraviolet wavelength, and its implications for young 
exoplanetary atmospheres \citep{guedel07}. 
When compared to other
targets in the `Sun in Time' sample, $\kappa$ Ceti is a moderately 
active star (spectral type G5 V) with a rotation period of 9.2 days
and an age of $\sim$750 Myrs \citep{guedel97}. Despite its
moderate rotation and activity, multi-year observations show a highly variable photosphere
with strong quasi-periodic photometric variability \citep{messina02}.
However, chromospheric activity measurements of the star
by the Mount Wilson project \citep{wilson78} reveal a more complex
variability with the presence of multiple chromospheric activity
cycles \citep{saar92,baliunas95,borosaikia18a}. 
This shows that $\kappa$ Ceti falls under a class of 
highly variable young Suns 
\citep{baliunas95,metcalfe13,olah16} whose long-term
variability could have a strong influence on any orbiting exoplanet's atmospheric 
evolution \citep{johnstone19,donascimento16}.
Hence, a detailed study of the temporal evolution of the 
star's magnetic field, wind, and high energy radiation can 
provide valuable information on how young Suns such as
$\kappa$ Ceti impact the development and evolution 
of habitable exoplanetary atmospheres.

Direct measurements of the surface magnetic field in Sun-like stars
depend on the Zeeman effect on polarised or unpolarised 
spectra. By measuring the Zeeman broadening 
in unpolarised spectra, 
both \citet{saar92} and \citet{kochukhov20} determined 
$\kappa$ Ceti's mean surface field strength to be $\sim$ 0.5 kG,
which includes contributions from both small- and 
large-scale magnetic features.
Using the technique of Zeeman Doppler 
imaging \citep[ZDI,][]{semel85,semel89,donati97,
kochukhov02,piskunov02,donati06,folsom18} 
on circularly polarised spectra of the star, an average vector 
magnetic field strength of $\geq$20 G was 
determined by \citet{rosen16} and \citet{donascimento16}, where the
small-scale features cancel out due to the use of circular 
polarisation and only the global large-scale field remains.
On both scales, the surface magnetic field of $\kappa$ Ceti is at least
a factor of 3-5 stronger than the solar surface magnetic field
\citep{kochukhov20,vidotto14}.
Using the ZDI magnetic field as input, simulations of its stellar wind 
\citep{donascimento16,airapetian21} reveal a highly energetic stellar environment 
with a high wind mass loss rate, which is about 
50-100 times stronger than the current solar wind. 

Long-term chromospheric activity monitoring 
of $\kappa$ Ceti reveals a 
highly irregular magnetic activity 
evolution \citep{baliunas95,hall07}. 
Lomb-Scargle period analysis 
\citep{lomb76,scargle82} on chromospheric 
activity measurements by \citet{baliunas95} showed the presence of a
5.6$\pm$0.1 year primary cycle period and a longer secondary 
cycle period. Re-analysis
of the same data set using a generalised Lomb-Scargle periodogram
\citep{zechmeister09} by \citet{borosaikia18a} shows that the 
secondary cycle period has a duration of 22.3 years. 

ZDI reconstructions of this star over two epochs 
by \citet{rosen16} 
and \citet{donascimento16} show that the star’s large-scale surface 
magnetic field is strongly toroidal and it changes geometry within a year. 
Such behaviour of the large-scale field is not observed in the Sun, but is now 
known to be common in rapidly rotating Sun-like stars \citep{petit08}.
Hence, in this work we carry out a detailed analysis of $\kappa$ Ceti’s magnetic field 
and activity over multiple epochs to understand how the star’s magnetic output changes over time. 
In a future work, we will investigate if and how the rapidly evolving large-scale 
magnetic field of $\kappa$ Ceti impacts the properties of its strong stellar wind.

This paper is organised as follows, in Section \ref{sec2} we 
briefly discuss the spectropolarimetric observations followed 
by the archival chromospheric measurements. Section \ref{sec3}
describes the two period analysis techniques used on the chromospheric data, and 
the technique of Zeeman Doppler imaging. Finally, 
the results are discussed in Section 4, and the conclusions 
are provided in Section 5.
%
\section{Observational data}\label{sec2}
\subsection{High resolution polarised and unpolarised spectra}
We obtained spectropolarimetric and 
spectroscopic data using the NARVAL instrument 
at the 2 m Telescope Bernard Lyot (TBL) at Pic du Midi \citep{auriere03}, 
some of which were observed as part of the BCool 
collaboration\footnote{https://bcool.irap.omp.eu} \citep{marsden14}.
NARVAL is a high-resolution spectrograph/spectropolarimeter
with a resolving power $R\sim$65,000, and a wavelength
range of 370 to 1050 nm. Observations were taken in
the spectropolarimetric mode, providing both circularly
polarised (Stokes {\it V}) and unpolarised (Stokes {\it I}) spectra.

The data were reduced using the 
standard Libre-ESpRIT reduction pipeline 
at TBL \citep{donatilsd}. The NARVAL spectropolarimetric 
data set consists of five observing epochs, 
2012.8, 2016.9, 2017.8, 2018.6, and 2018.9 
spanning over six years. 
In addition to the NARVAL data we included 
epoch 2013.7 from \citet{rosen16} which was obtained using the 
HARPSpol spectropolarimeter at 
the 3.6 m ESO telescope at La Silla, Chile \citep{snik11,piskunov11}. 
HARPSpol has a wavelength coverage of 360–691 nm~
and a resolving power of $\sim$ 110,000. 
Although epoch 2012.8 and 2013.7 were previously
investigated by \citet{rosen16} and \citet{donascimento16}, we reanalysed
them for completeness and comparison purposes.
Table \ref{journal} lists the individual observations taken over 
our entire spectropolarimetric data set of six epochs. 
The reduced NARVAL spectra are also available at the PolarBase archive
\citep{petit14}.

\subsection{Archival S-index measurements}
In addition to high-resolution spectra from NARVAL/TBL and HARPSpol
we obtained chromospheric activity measurements from three other
sources, Mount Wilson \citep{wilson78}, 
Lowell \citep{hall95}, and TIGRE \citep{schmitt}. 
Chromospheric activity in cool stars is often determined 
by measuring the S-index of the star, which is the flux in the 
\ion{Ca} {II} H and K lines normalised
to the nearby continuum, as described in \citet{duncan91}. The term 
S-index ($S_\mathrm{MWO}$) was first coined by the Mount Wilson project 
\citep{wilson78,duncan91,baliunas95}, where stellar 
$S_\mathrm{MWO}$ measurements were carried out between 1966-2002.
The Mount Wilson $S_\mathrm{MWO}$ measurements used in this work, 
shown in Fig.~\ref{appendixS}, were
taken from the publicly available `1995 compilation' of the Mount Wilson HK
project released by the National Solar Observatory (NSO)
\footnote{https://nso.edu/data/historical-data/mount-wilson-observatory-hk-project/}. 

S-index measurements of $\sim$100 Sun-like stars were also carried out
by the Solar-Stellar Spectrograph (SSS) at the Lowell observatory
as part of a long-term monitoring programme. As part of this programme 
$\kappa$ Ceti was regularly observed between 1993 and 2018. The Lowell S-indices used in this
work are calibrated $S_\mathrm{MWO}$ measurements \citep{hall95}, shown in   
Fig.~\ref{appendixS}.

Since 2014, $\kappa$~Ceti has also been continuously monitored by the 
TIGRE facility at the La Luz Observatory in Mexico, a 1.2~m telescope connected 
to the HEROS spectrograph with a spectral resolution of 20,000 
and a wavelength range of 380-880~nm. TIGRE is a dedicated
instrument for the study of stellar magnetic activity. 
The TIGRE data of $\kappa$~Ceti have previously 
been used by \citet{hempelmann16} and \citet{schmitt2020} for 
the determination of the star's rotation period. 
The TIGRE S-index measurements (see Fig.~\ref{appendixS}) were calibrated 
to $S_{\mathrm{MWO}}$ via the conversion derived by \cite{mittag16}.

\section{Methods and data analysis}\label{sec3}
%
\subsection{Time series analysis of $S_\mathrm{MWO}$}\label{wvletlabel}
\begin{figure}
\includegraphics[width=1.\columnwidth]{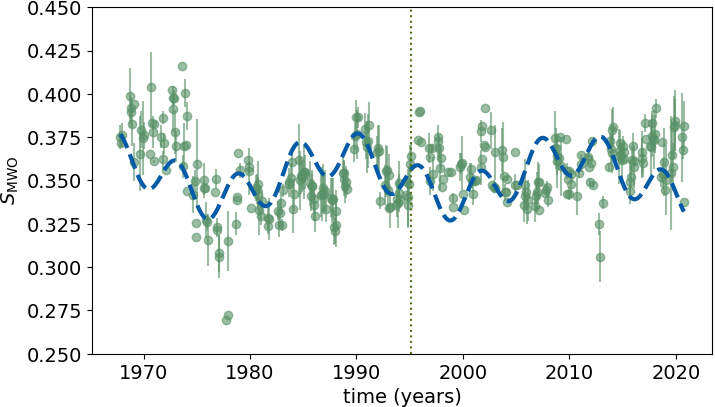}
\caption{S-index ($S_\mathrm{MWO})$ vs time. Monthly averaged values 
are plotted as green circles, where the dispersion is shown as error bars.
The dotted vertical line separates the $S_\mathrm{MWO}$ from 
this work (to the right of the vertical line) and \citet{borosaikia18a} (to the left of the 
vertical line). The best fit cycle periods (22.3 and 5.7 years) from 
\citet{borosaikia18a} are shown in blue.} 
\label{fig1}
\end{figure}

We measured the 
S-index of $\kappa$ Ceti using spectroscopic data 
taken at NARVAL/TBL, and calibrated our measurements to the historical
Mount Wilson S-index, $S_\mathrm{MWO}$, using the 
coefficients determined in  \citet{marsden14} \citep[also see][for more details on the 
S-index measurements]{borosaikia16}. Our new $S_\mathrm{MWO}$ 
measurements when combined with existing data lead to a time series
baseline of more than 50 years, as shown in Fig.~\ref{fig1}. 
To characterise the temporal evolution of the 
$S_\mathrm{MWO}$ measurements,  we applied a 
wavelet decomposition \citep{torrence98} to the time series, in addition to a Generalised Lomb Scargle (GLS) 
periodogram \citep{zechmeister09}. 

The GLS periodogram is a state-of-the-art algorithm that is well suited for detecting 
periodic signals in an unevenly sampled time series.
The algorithm can be compared to a least-squares fitting of sine functions, and is conceptually similar to 
the Lomb-Scargle periodogram of \citet{lomb76} and \citet{scargle82}. The key difference is the use of an offset and weights 
based on the measurement errors, which are not implemented 
in the original Lomb-Scargle algorithm. 
The version of GLS used in this work was developed by 
\citet{zechmeister09}, and we pre-processed our activity time series
by binning the data into monthly bins before applying the GLS algorithm, 
as shown in Fig.~\ref{fig1}, where the dispersion in the data 
is shown as error bars. To calculate the false alarm probability 
(FAP) of the detected signals we applied the normalisation of \citet{horne86}
and the method described in \citet{zechmeister09}.      
%

One key disadvantage of the GLS algorithm or any other Fourier method is their ineffectiveness towards non-stationary 
signals, that is they are strong in period detection 
but do not provide temporal information of any detected periods.  
Additionally, in the case of a irregular non-sinusoidal period
the harmonics might dominate over the true period.
Wavelet analysis, on the other hand, can determine both the periodicities present in the signal and their duration 
and localisation in the time series. The specific wavelet analysis code used in this work is a modification 
of the one described in \citet{torrence98}. A Morlet function is used on a set of temporal scales and 
the wavelet is calculated for each scale, as well as the `cone of influence' (COI) – the region of the 
wavelet spectrum in which edge effects (due to the fact that the time series are finite) become important. 
As discussed in \citet{torrence98}, the wavelet transform can be considered as a band-pass filter with a known response 
function, which filters the time series by summing over a subset of scales and also reconstructs 
the filtered signal. Here, we have chosen to use scales of less than 7 years, since most of the higher periodicities 
(even if real) will fall in the COI region (see Section 4.1). 
For determining the statistical significance of the filtered data and the probability level, we used the randomisation 
method described in detail in \citet{oshea01}.
\subsection{Least squares deconvolution and Zeeman Doppler imaging}
%
While $S_\mathrm{MWO}$ is a well studied magnetic activity 
proxy, detection of polarity reversals in the vector 
magnetic field is essential to constrain the star's dynamo 
operated magnetic cycle \citep{petit09,fares09,borosaikia16,mengel16,jeffers18}.
We used the spectropolarimetric observations 
to determine the star's large-scale magnetic field strength and 
geometry. We applied the technique of least squares 
deconvolution \citep[LSD,][]{donatilsd,kochukhov10} to boost the S/N of the observations,
and used ZDI \citep{donati06,folsom18}
to reconstruct the large-scale surface magnetic field of $\kappa$ Ceti. 

\begin{figure*}
\centering
\includegraphics[width=0.78\textwidth]{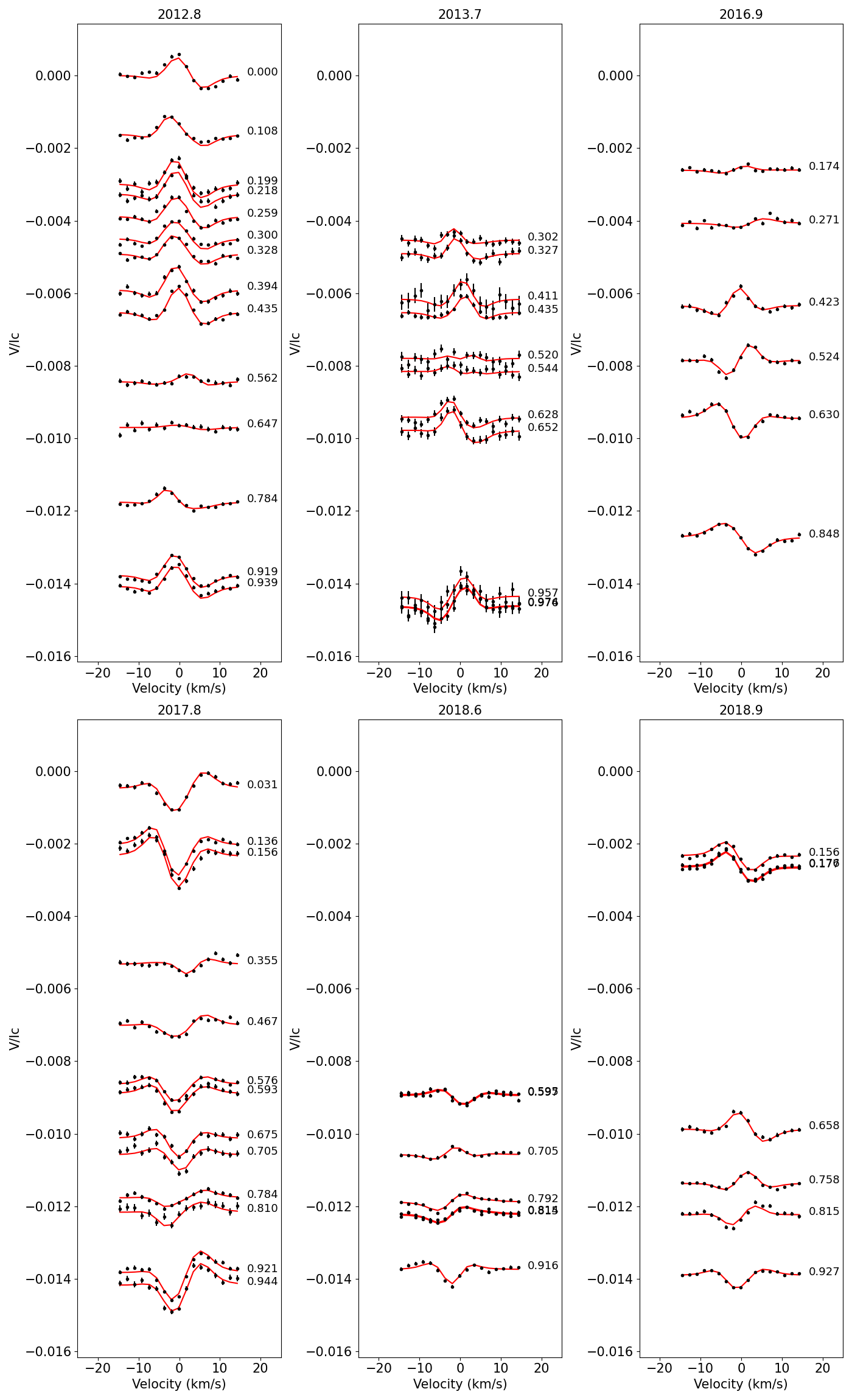}
\caption{Observed (black) and modelled (red) Stokes~{\it V} LSD profiles for the six
epochs of observations. Rotational phases are shown next to each profile. The profiles
are shifted vertically based on their rotational phases, where the phases
increase from 0 to 1 (top to bottom). }
\label{lsdfig}
\end{figure*}

In Sun-like cool stars magnetic field signatures are extremely hard to detect in 
individual polarised spectral lines. Hence, we used the multi-line technique of 
LSD on our circularly polarised Stokes {\it V} and unpolarised 
Stokes {\it I} spectra \citep{donati97,kochukhov10}. 
To obtain an averaged LSD line profile
we assumed that all magnetically sensitive lines in the observed spectra have the 
same characteristic shape. 
The final averaged LSD spectral line
profiles were created by deconvolving the observed stellar spectra with a line mask. 
The line mask used in this work was taken from 
\citet{marsden14}, which corresponds to 
a $T_\mathrm{eff}$ of 5750 K and a $\log g$ of 4.5. 
The mask was created from the 
Vienna Atomic Line Database \citep[VALD,][]{kupka000} using all
lines that had a depth $\geq$ 10 percent of the continuum.
Exceptionally broad lines such as Balmer 
lines were excluded from the mask. 
The normalisation parameters used 
for the LSD profiles were, a line depth of 0.52, a Land\'e factor of
1.22, and a central wavelength of 580 nm. 
The LSD (Stokes {\it V}) profiles for each epoch, 
containing a time series of 6-14 observations, are shown in Fig.~\ref{lsdfig} in black. 

The tomographic technique of ZDI is an inverse routine 
that inverts observed LSD line profiles to 
reconstruct the large-scale surface magnetic geometry of 
stars. 
The version of ZDI used in this work \citep{folsom18} uses a maximum entropy fitting, where model 
LSD line profiles are iteratively fit to observed LSD line profiles. 
The model LSD profiles  for any given epoch were created 
using a spherical stellar model of $\kappa$ Ceti. 
All model stellar properties, except the radial velocity, were 
taken from the literature, as shown in Table
\ref{tableproperties}. A simple Gaussian fit was 
performed on the observed LSD Stokes {\it I} profiles to 
determine the radial velocity of the star. 

\begin{table}
\centering
\caption{Stellar properties of $\kappa$ Ceti.}
\label{tableproperties}
\begin{tabular}{lcc}
\hline
\hline
Parameter&Value&Reference\\
\hline
Mass (M$_\sun$)& 0.948& 1\\
Radius (R$_\sun$)&0.917 & 1\\
Radial velocity, $v_\mathrm{r}$ (kms$^{-1}$)&19.17&this work\\
Rotational velocity, $v\sin i$ (kms$^{-1}$)& 5.2&1\\
Rotation period, $P_\mathrm{rot}$ (days)&9.2& 2\\
Inclination ($^\mathrm o$)&60& 2\\
Age (Myrs)&750& 3\\
\hline
\end{tabular}
\tablebib{(1)~\citet{valenti05}; (2) \citet{rosen16}; (3) \citet{guedel97}
}
\end{table}
To create the model LSD line profiles our stellar model was
divided into surface elements of equal area. Each surface element was
assigned a local model line profile. The local model Stokes {\it I} 
lines are approximated as Voigt profiles and the local model
Stokes {\it V} lines are created under the weak field approximation
\citep{landi04}. The model line
profiles depend on the line-of-sight magnetic field strength,
which in turn was calculated from a set of 
complex valued spherical harmonics coefficients ($\alpha_\mathrm{lm}$, $\beta_\mathrm{lm}$, 
and $\gamma_\mathrm{lm}$, where $l$ is the spherical harmonics degree and
$m$ is the spherical harmonics order) that
describe the magnetic field \citep{donati06}. 
The set of spherical harmonics were limited to a maximum
degree $l_\mathrm{max}$ of 10 to be consistent with previous
work by \citet{rosen16} and \citet{donascimento16}. Additionally, due to the 
low $v \sin i$ of the star increasing $l_\mathrm{max}$ to 
include higher degrees does not have any significant impact 
in the magnetic field reconstructions.
The spherical harmonics
coefficients are the free parameters during the fitting
process and they define the three magnetic field vectors, radial 
$B_\mathrm{r}$, meridional $B_\mathrm{\theta}$, and azimuthal 
$B_\mathrm{\phi}$ \citep[see][for more details]{donati06,folsom18}.
The final model LSD line profiles for a given observation were created by summing 
all the local line profiles from the visible 
surface area and scaled by a linear limb darkening law
with a coefficient of 0.66. The final model LSD profiles were 
iteratively fit to the observed LSD profiles using the 
maximum entropy method \citep{skilling84}. Figure \ref{lsdfig}
shows the best fit model LSD profiles in red.

The maximum entropy method is a regularisation routine that minimises the 
$\chi^2$ and maximises the entropy, and is an essential step for inverse problems such as ZDI. 
In this regularisation routine the 
user assigns a target $\chi^2$ known as $\chi^2_\mathrm{aim}$
that the routine aims to achieve by minimising $\chi^2$. Once $\chi^2=\chi^2_\mathrm{aim}$ 
is achieved the routine maximises the entropy till the maximum 
possible entropy is reached for $\chi^2=\chi^2_\mathrm{aim}$.
The entropy definition is applied to the spherical 
harmonics coefficients $\alpha_\mathrm{l,m}$, $\beta_\mathrm{l,m}$, and 
$\gamma_\mathrm{l,m}$ which are the free parameters
of the model.
Table~\ref{chi2table} lists the reduced $\chi^2$ achieved for the six epochs of observations.
We do not include differential rotation in the ZDI reconstructions of the
large-scale field, which could explain the slightly higher values 
of the reduced $\chi^2$.
For a more detailed description of the ZDI technique and the regularisation routine 
please refer to Appendix B in \citet{folsom18}.
%
\begin{figure}
\centering
\includegraphics[width=.5\textwidth]{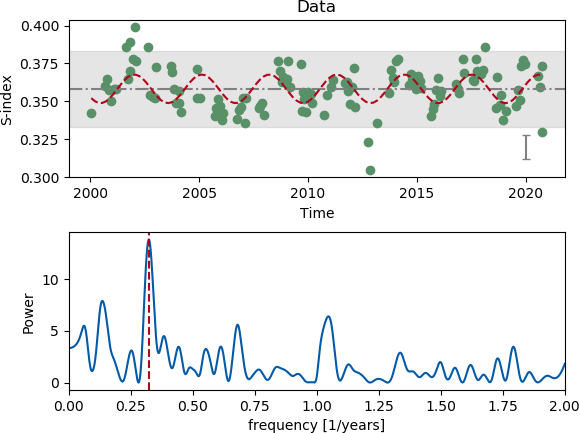}\\
\caption{S-index of $\kappa$ Ceti over a 20 year time period in the top plot.
The grey band marks the range of the solar S-index variability during a solar cycle.
The 3.1 year cycle period is shown as a red dashed curve.
The periodogram is shown in the bottom plot, where the strongest frequency is 
marked by a dashed red vertical line.}
\label{gls}
\end{figure}

%
\section{Results and discussion}\label{sec4}
%
\subsection{Chromospheric activity cycles of $\kappa$ Ceti}\label{chromcycle}
%
The chromospheric activity ($S_\mathrm{MWO}$) of $\kappa$ Ceti 
is highly variable and exhibits multiple periodicities, as 
shown in Fig.~\ref{fig1}. In a previous study
we identified two chromospheric activity cycles with a period of 22.3 years and 5.7 years
using the GLS periodogram \citep{borosaikia18a}. The blue dashed line in Fig.~\ref{fig1}
represents the two activity cycles, and they are in good agreement with 
the observed $S_\mathrm{MWO}$ measurements
from the late 1970s to the early 2000s. However, the agreement between the
two detected cycle periods and the $S_\mathrm{MWO}$ measurements is weaker 
in the very early parts and the later half of the time series, as shown in 
Fig.~\ref{fig1}. This could be either due
to variable duration of one or both cycle periods, 
complex temporal evolution of the cycles, 
presence of multiple other periodicities that dominate 
in the beginning and the later part of the time series, or any combination of
the above.

To investigate this further, we applied the GLS periodogram on 
the new observations (2000-2020), which is a sub-set of the time series in
Fig~\ref{fig1}. The strongest peak of the periodogram lies at 
3.1$\pm$0.01 years, as shown in Fig.~\ref{gls}. 
The FAP of this cycle period is $1.3\mathrm{e-}05$,
which is not as robust as the 5.7 year (FAP = $6.3\mathrm{e-}51$) cycle period in \citet{borosaikia18a}.
While the 3.1 year could be a harmonic of the 5.7 ($\sim$ 6) year period,
we could not detect the 22.3 year period which could be due to the
shorter time series of 20 years. On closer inspection of Fig.~\ref{fig1},
the overall trend in the later part of the time series (2000 onwards) appears
to be more flat compared to the pre-2000 time series, which could also contribute to
the lack of detection of the longer 22.3 year period.

\begin{figure}
\centering
\includegraphics[width=.52\textwidth]{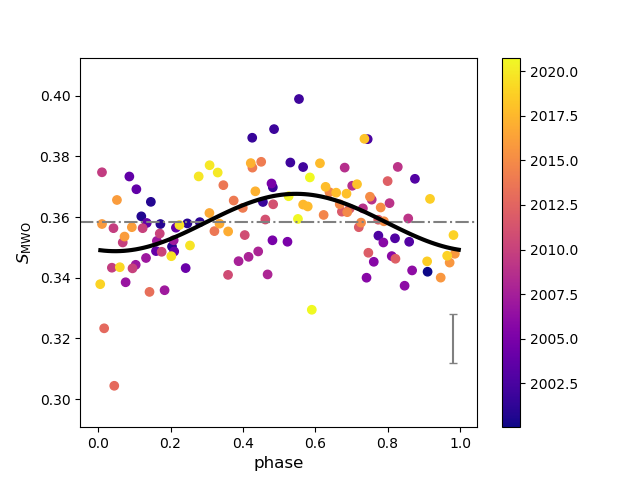}
\caption{Same S-indices as Fig.~\ref{gls}, but the data were phase-folded using
a cycle period of 3.1 years (in black). The colour bar represents the observational
time-span in years.}
\label{phasefolded}
\end{figure}

The period analysis on the shorter data set (2000-2020) in Fig.~\ref{gls} shows that 
the dominant cycle period has a duration of 3.1 years. However, 
this period on its own does not explain the complex 
variability in the data set. This is clearly seen in the phase-folded S-indices in 
Fig.~\ref{phasefolded}, where the data shows a weak agreement with the 3.1 year
cycle period. The complexity of $\kappa$ Ceti's
S-index variability lies far beyond the cycle periods identified by the GLS
periodogram and information on the temporal evolution of its cycle periods is 
needed.

\begin{figure}
\centering
\includegraphics[width=1\columnwidth]{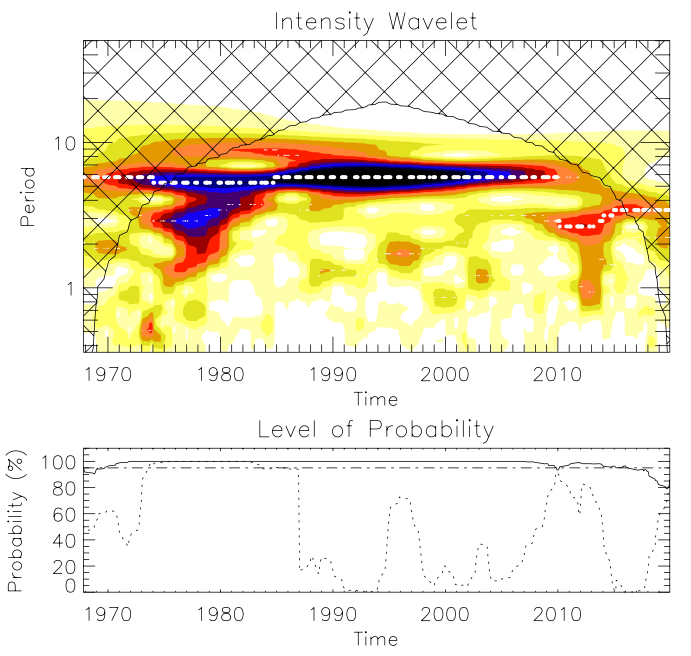}
\caption{{\it Top}: Wavelet intensity spectrum of monthly averaged $S_\mathrm{MWO}$ of
$\kappa$ Ceti. The strongest cycle periods are represented by the darker 
colours and the dashed horizontal white lines, and the cross-hatched area shows the COI.
{\it Bottom}: The probability associated with the two detected periods of 5.8 
(solid black line) and 3.1 (dotted black line) years. The 90$\%$ probability level is
shown as a horizontal dashed and dotted line. }
\label{waveletfig}
\end{figure}

We applied a wavelet transform to $\kappa$ Ceti's monthly 
averaged $S_\mathrm{MWO}$ measurements, as described in Section~\ref{wvletlabel}. 
Our wavelet code detected two dominant cycle periods of 
5.8 years and 3.1 years including their occurrence in time, with a probability 
level of $99\% - 100\%$ for the two detected periods, as shown in Fig.~\ref{waveletfig}. 
The 22.3 years cycle detected using the GLS periodogram
in \citet{borosaikia18a} falls inside the COI of our transform and
is not included in this work. The COI, shown as a cross-hatched area in 
Fig.~\ref{waveletfig}, marks the region within which the period determination is
unreliable, due to edge-effects and limited time series length \citep{torrence98}.
The 5.8 cycle period in Fig.~\ref{waveletfig} is in good agreement with the 5.7 year 
cycle period determined from monthly averaged measurements in 
\citet{borosaikia18a}, and it is the only dominant cycle period 
between early 1980's and mid 2000's, as shown in Fig.~\ref{waveletfig}. 
In the first $\sim$10 years of the time series, the 
3.1 year cycle co-exists with the 5.8 year cycle, 
where as the 3.1 year cycle is the 
only dominant cycle period from $\sim$2008 onwards.
A wavelet decomposition on the 2000-2020 data set also shows
that the 3.1 year is the dominant cycle period during this
period (Fig.~\ref{newave}).   

Over a time period of 
50 years the chromospheric activity of $\kappa$ Ceti exhibits 
two strong periodicities
with variable temporal evolution. 
The $\sim$1:2 ratio
between the two periods suggest that the 3.1 year period is
the first harmonic. 
This shows the importance of 
long-term monitoring of stellar
activity data, specially for active young Suns with
highly irregular chromospheric activity. 
Based on when the star is observed 
one would only detect the 3.1 year cycle period, 
with a non-detection of the 5.7-5.8 (referred as the
5.7 from here onwards) year period, as shown
in Figures~\ref{gls} and \ref{newave}. 

Multiple cycle periods have been also detected in
a limited number of Sun-like stars \citep{baliunas95,olah16}. 
As an example, a wavelet decomposition on the active
exoplanet-host star $\epsilon$ Eridani by \citet{metcalfe13} 
suggests a possible $\kappa$ Ceti-like behaviour of the star's two dominant cycles. 
Recent multi-wavelength 
observations of $\epsilon$ Eridani by \citet{coffaro20} and 
\citet{petit21} indicates that the star's 
magnetic activity is approaching a more chaotic regime.
Period analysis of long-term stellar activity measurements 
will help us determine if such complex evolution of multiple 
activity cycles is indeed a common occurrence in young Suns. 

\begin{table}
\centering
\caption{\label{chi2table} Reduced $\chi^2$ of the ZDI magnetic maps.}
\begin{tabular}{cc}
\hline
\hline
epoch&$\chi^2$\\
\hline
2012.8&1.5\\
2013.7&1.0\\
2016.9&1.1\\
2017.8&1.3\\
2018.6&1.1\\
2018.9&1.2\\
\hline
\end{tabular}
\end{table}

Even a moderately active cool star such as our Sun exhibits 
multiple periodicities in its activity and magnetic field measurements.
The most prominent and well studied solar cycle is the 
sunspot cycle, also known as the Schwabe cycle, which 
exhibits a periodicity of 11${^{+3}_{-2}}$ years \citep{usoskin17}.
The 11 year Schwabe cycle is closely tied to the 22 year magnetic cycle 
or Hale cycle, where the global magnetic field 
of the Sun flips its polarity. The Hale and Schwabe
cycles have a 1:2 ratio, and are tied to the underlying tachocline dynamo of the Sun. 
On the surface $\kappa$ Ceti's 3.1 and 5.7 year cycles also exhibit a $\sim$1:2 ratio
similar to the two dominant solar cycles.  
However, $\kappa$ Ceti's cycles exhibit a far more complex behaviour.
While the 5.7 year cycle
dominates for the majority of the time series, it 
gets considerably weaker in the later part of the time series, as 
shown in Fig.\ref{waveletfig}. This 
is a stark difference from the behaviour of the two dominant
solar cycles. Furthermore, for reliable comparison
with the solar Hale and Schwabe cycles one must have
information on $\kappa$ Ceti's magnetic field polarity flips.

Based on the chromospheric activity data alone the complex evolution of 
$\kappa$ Ceti's two strong cycle periods could be
attributed to strong stellar 
surface inhomogeneities. As an example, the Sun 
also exhibits multiple other periodicities apart from the 
two dominant cycles mentioned above. 
The most prominent of which is the {\it Gleissberg} 
cycle of $\sim$90 years \citep{gleissberg39}, which exhibits 
significant variations in amplitude and duration. The {\it Gleissberg} cycle is
associated with sunspot appearance and is shown to have existed for at least within
the last millennia \citep{ogurtsov02}, although it is known to sometimes completely 
disappear \citep{beer18}. Investigations of the north-south asymmetries of magnetic 
activity in the Sun have also resulted in the detection of a cycle period 
similar to the {\it Gleissberg} cycle and multiple other short term cycle 
periods \citep{deng16}. However, the amplitudes of the 
multiple periods detected in $\kappa$ Ceti are much stronger 
than in the multiple periods detected in the solar case. 
Hence additional measurements of the photospheric vector
magnetic field of $\kappa$ Ceti is crucial to reliably 
characterise the two cycles in Fig.~\ref{waveletfig}. 
\begin{figure*}[h]
\centering
\begin{subfigure}[t]{0.3\textwidth}
\centering
\includegraphics[width=1\textwidth]{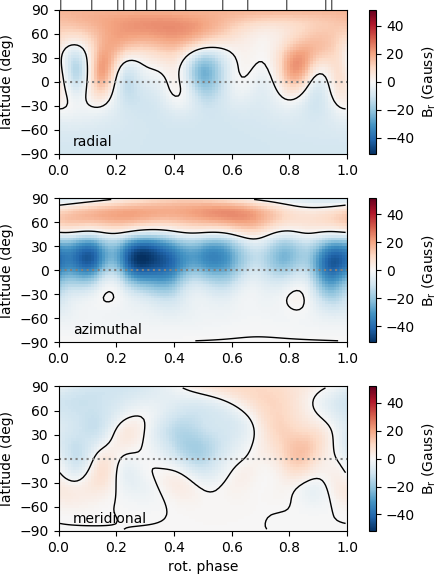}
\caption{2012.8}
\end{subfigure}
\begin{subfigure}[t]{0.3\textwidth}
\centering
\includegraphics[width=1\textwidth]{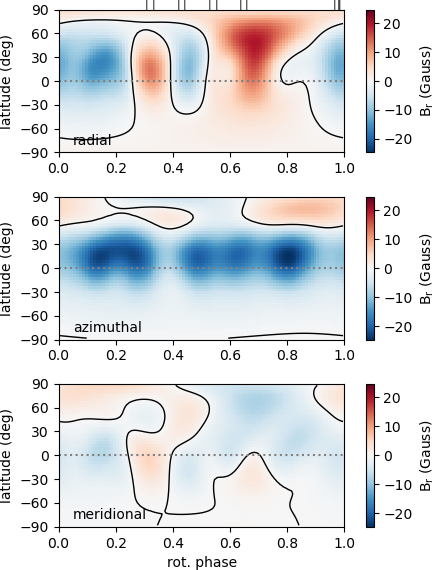}
\caption{2013.7}
\end{subfigure}
\begin{subfigure}[t]{0.3\textwidth}
\centering
\includegraphics[width=1\textwidth]{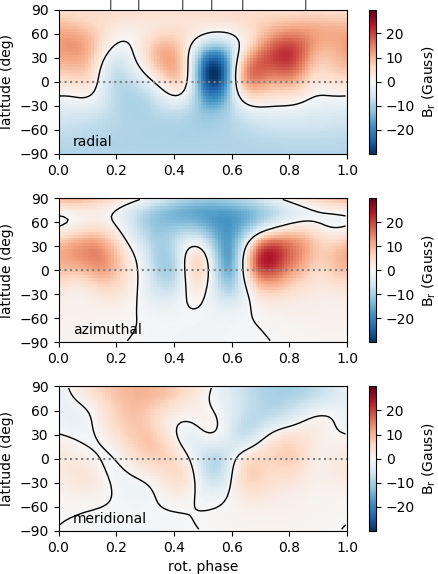}
\caption{2016.9}
\end{subfigure}
\par\bigskip
\begin{subfigure}[t]{.3\textwidth}
\centering
\includegraphics[width=1\textwidth]{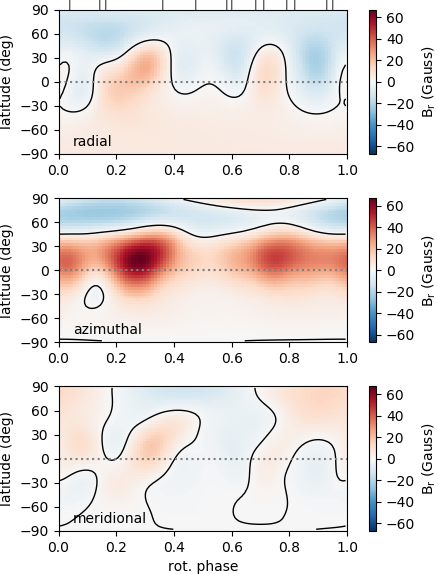}
\caption{2017.8}
\end{subfigure}
\begin{subfigure}[t]{.3\textwidth}
\centering
\includegraphics[width=1\textwidth]{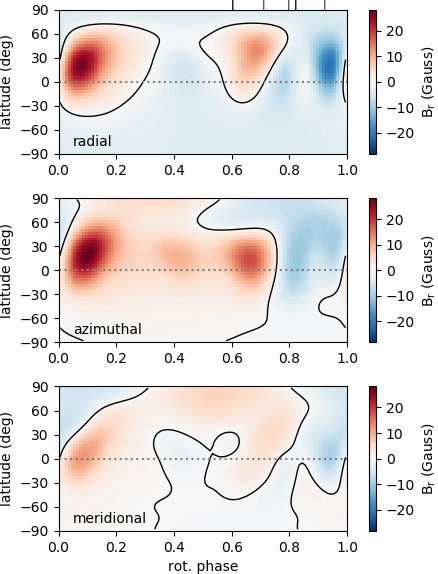}
\caption{2018.6}
\end{subfigure}
\begin{subfigure}[t]{.3\textwidth}
\centering
\includegraphics[width=1\textwidth]{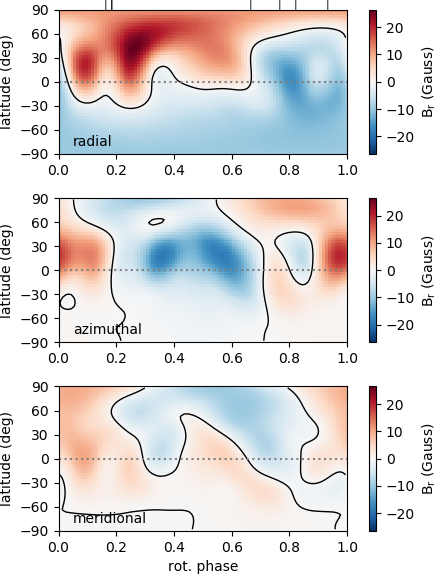}\\
\caption{2018.9}
\end{subfigure}
\caption{Surface magnetic maps of $\kappa$~Ceti for six 
observational epochs. The large-scale radial, azimuthal 
and meridional field is shown in each sub-plot. 
The rotational phase is shown in the x-axis and the 
latitude is shown in the y-axis. The equator is marked by a horizontal dotted
line. The magnetic field strength is shown in gauss, 
where red denotes positive polarity and blue denotes negative polarity. 
The black tick marks at the top mark the observed rotational phases. 
The colour scale is determined based on the maximum magnetic field
strength in each ZDI map.}
\label{maps}
\end{figure*}
\subsection{Magnetic field geometry and polarity flips}
Reconstructions of the large-scale surface magnetic field
geometry were carried out using the technique of ZDI for
six observational epochs.
Figure~\ref{maps} shows the 
ZDI surface magnetic maps for the vector magnetic field 
in the three directions, radial, azimuthal, and meridional.
The meridional field is much weaker 
than the radial and the azimuthal field, as shown 
in Fig.~\ref{maps} (bottom panels). It has
been shown by previous studies that possible cross-talk between 
the radial and meridional components lead to underestimation 
of the latter \citep{donati97,rosen15,lehmann19}. 
Hence we only discuss the properties of
the radial and the azimuthal field.
By definition the radial component of the magnetic field 
relates to the poloidal magnetic field and the 
azimuthal and meridional components of the magnetic field relate to 
both poloidal and toroidal component of the magnetic field
\citep[See ][for more details]{donati06,folsom18}. 

\subsubsection{Radial field}
The radial field of $\kappa$ Ceti 
exhibits considerable evolution over a six 
year period, as shown in Fig.~\ref{maps} (top panels). A positive polarity radial magnetic 
field dominates the northern hemisphere of the star at epoch 2012.8. 
At epoch 2013.7 and 2016.9 the positive polarity is not as dominant 
as in epoch 2012.8 and the northern hemisphere 
consists of both strong positive and negative 
magnetic field. A negative polarity magnetic field 
takes over at epoch 2017.8, indicating a polarity 
flip from positive to negative. 
The radial field undergoes a reconfiguration a year later 
at epoch 2018.6 exhibiting a mixed polarity in the northern hemisphere. 
At epoch 2018.9 the mixed polarity evolves to show a 
stronger positive polarity magnetic field. 
Although the sparse phase coverage of 
epochs 2018.6 and 2018.9 make it difficult to obtain 
robust magnetic maps, the Stokes {\it V} profiles in
Fig~\ref{lsdfig} show considerable variation in the amplitude 
and signature, providing confidence that the magnetic
field undergoes a rapid evolution towards the end of our observational
time-span.

The evolution of the radial magnetic field 
component shows strong evidence of 
a polarity flip from 2012.8 to 2017.8, 
indicating it could take 5 years for 
the star's large-scale radial magnetic 
field to flip polarity. In those five years
(between 2012.8 and 2017.8) only two other (2013.7 and 2016.9)
magnetic maps are available. 
At epoch 2013.7 the radial field exhibits both 
positive and negative polarity regions, indicating a much more
complex field than a year earlier at epoch 2012.8. 
Epoch 2016.9 also shows a complex magnetic field, with 
the presence of both negative and 
positive polarities in the 
northern hemisphere, so the only other time
the star could have undergone a polarity switch from a 
positive to a negative magnetic field before 2017.8 is 
between 2013.7 and 2016.9.
This will suggest that the dominant positive magnetic field
in 2012.8 flips to a dominant negative field between
2013.7 and 2016.9, and turns back to positive in 
2017.8, resulting in a magnetic or Hale cycle
of $\sim$ 6 years. 
However, Fig~\ref{maps} shows that at epoch 2017.8 the radial field is 
dominantly negative not positive as expected, indicating that the $\sim$ 6 year Hale
cycle is not feasible.

Based on our observations, 
the flip in 2017.8 is most likely the first 
polarity flip between epoch 2012.8 and 2017.8, and we expect
the Hale cycle to have a period of $\sim$10 years. 
Since a majority of our observations were taken 
at yearly epochs any polarity flips that might have occurred at 
monthly intervals were missed. Future 
high cadence observations will help us fully characterise 
the Hale cycle.

\subsubsection{Azimuthal field} 
The azimuthal field evolves from a simple almost single polarity
field to a more complex field during the course of our 
observations, as shown in Fig.~\ref{maps} (middle panels). It exhibits a 
band of strong negative polarity field at equatorial 
latitudes at epochs 2012.8 and 2013.7. 
Over time the field evolves, 
with the appearance of a positive polarity magnetic field at epoch 
2016.9, which switches to a dominant positive polarity 
magnetic field at epoch 2017.8. The dominant positive polarity 
azimuthal field re-configures at epoch 2018.6, 
with the appearance of negative polarity regions. 
Epoch 2018.9 shows the presence of both 
positive and negative large-scale magnetic field. 
\begin{table*}
\centering
\caption{\label{Btable}Magnetic properties.}
\begin{tabular}{ccccccc}
\hline
\hline
epoch&B$_\mathrm{mean}$ (G)&pol ($\%$total)&dipole ($\%$poloidal)&quad ($\%$poloidal)&oct ($\%$poloidal)&axi ($\%$total)\\
\hline
2012.8&21$\pm$2&30$\pm$5&44$\pm$8&17$\pm$3&15$\pm$3&74$\pm$2\\
2013.7&11$\pm$1&36$\pm$1&37$\pm$4&27$\pm$1&22$\pm$2&64$\pm$2\\
2016.9&12$\pm$1&85$\pm$3&51$\pm$4&14$\pm$2&23$\pm$3&25$\pm$3\\
2017.8&20$\pm$1&26$\pm$1&24$\pm$5&43$\pm$4&13$\pm1$&74$\pm$1\\
2018.6&9$\pm$1&77$\pm$1&18$\pm$1&43$\pm$2&20$\pm$2&19$\pm$2\\
2018.9&12$\pm$1&81$\pm$3&71$\pm$2&12$\pm$3&5$\pm$1&25$\pm$3\\
\hline
\end{tabular}
\tablefoot {The mean magnetic field in each epoch is
followed by the fraction of the total (poloidal+toroidal) 
magnetic energy reconstructed
in the poloidal field component, the fraction of the total poloidal
field reconstructed in the dipolar, quadrupolar, and octupolar components, 
and the fraction of the total magnetic energy stored in the 
axisymmetric component. \par}
\end{table*}

While the single polarity positive radial field changes to a more complex field from 
2012.8 to 2013.7, the azimuthal field primarily remains at a single 
polarity from 2012.8 to 2013.7, as shown in 
Fig.~\ref{maps}.  This discrepancy between the radial and the azimuthal field
could be explained by a time lag between the azimuthal field and the radial field. 
A time lag of 1-3 years is known to 
exist between the solar poloidal and toroidal magnetic fields \citep{cameron18}.
A time lag between the radial and the azimuthal field is also detected in the case of 
other Sun-like stars \citep{jeffers18,borosaikia18b}.
Taking this time lag into account, the azimuthal field
is in good agreement with the radial field suggesting it also
undergoes a polarity reversal. However this time lag is not detected from 2017.8 to 2018.6,
which suggests our epochs are too sparse to fully explore the 
complexity of $\kappa$ Ceti's magnetic field evolution. 
%
%
%
\subsection{Magnetic morphology}
The large-scale ZDI magnetic geometry of $\kappa$~Ceti is composed 
of both poloidal and toroidal field components, with the poloidal 
field dominating in epochs 2016.9, 2018.6, and 2018.9, and the
toroidal field dominating in epochs 2012.8, 2013.7, and 2017.8, as shown in Table~\ref{Btable}. 
The strong toroidal field corresponds to the appearance 
of the single polarity equatorial azimuthal field at 
epochs 2012.8, 2013.7, and 2017.8 in Fig.~\ref{maps}. It is not surprising as
young cool dwarfs like $\kappa$ Ceti are known to switch from 
dominantly poloidal to toroidal configuration over multi-epoch
observations \citep{petit08,borosaikia15,rosen16,see15}. Although, exceptions 
exists as the young solar analogue $\epsilon$ Eridani exhibits a consistently 
strong poloidal field over multiple epochs with a stronger toroidal field in 
only one out of nine epochs \citep{jeffers14,jeffers17}. Whereas, a dominantly
toroidal field is detected in the young Sun EK Dra over five epochs \citep{waite17}.
Our Sun and older cool stars like 61 Cyg A \citep{borosaikia16} 
exhibit a dominant surface poloidal field.

\begin{figure}
\centering
\includegraphics[width=.5\textwidth]{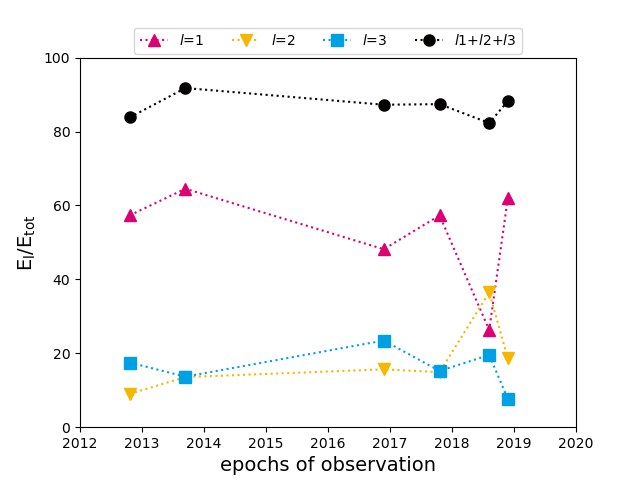}
\caption{Magnetic energies (poloidal+toroidal) 
for the three smallest values of spherical harmonic degree $l$.}
\label{l}
\end{figure}

The percentage of magnetic energies distributed 
between the different components of the star's poloidal 
magnetic field is shown in Table~\ref{Btable}. 
The poloidal field is strongly dipolar ($l$=1) at certain epochs, whereas
the quadrupolar ($l$=2) field dominates at other epochs.
Since the large-scale field is made up of both
poloidal and toroidal components
we also investigate the fraction of magnetic energy in 
the lower spherical harmonics order of the poloidal and 
toroidal field ($l$ = 1, 2, 3). Figure~\ref{l} shows that at all 
observed epochs $\geq$80$\%$ of the total energy (combined 
poloidal and toroidal) can be found in the lower order 
spherical harmonics degree $l$=1, 2 and 3. 

No clear periodicity is detected in
the fluctuations in the magnetic energies associated with the 
lower order harmonics, as shown in Fig.~\ref{l}. According to \citet{lehmann21} 
periodicities are not easily detectable in the magnetic 
energy fractions of ZDI magnetic maps. Instead, the 
authors identified the axisymmetric fraction as a good tracer of
the magnetic cycles. Although the axisymmetric fractions 
do not show any clear periodic behaviour, as shown in Table 
\ref{Btable} a weak anti-correlation
between its strength and the percentage of the 
poloidal energy is detected.

Similar to the analysis of \citet{rosen16} and \citet{donascimento16}, our 
results show that the majority of the magnetic energy
is concentrated in the lower harmonic degrees. 
However, slight discrepancies between the magnetic
properties derived in this work and \citet{rosen16} are also
detected, specifically for the HARPSpol data at epoch 2013.7. 
At epoch 2013.7
the magnetic field strength determined in this work is a factor of
2 weaker than the field strength obtained by \citet{rosen16}.
This discrepancy could be attributed to differences in 
the line mask used for LSD, 
the stellar line models, and the 
definition of the maximum entropy used in the ZDI code. Such
differences might be of greater importance 
for the high-resolution spectropolarimetric
data taken by HARPSpol, as no such discrepancy is detected 
in the 2012.8 data. 
Despite the differences in the mean magnetic 
field strength at epoch 2013.7 our reconstructed magnetic geometry in 
Fig.~\ref{maps} is in strong agreement with the magnetic
map in \citet{rosen16}. Hence, we are confident that the large-scale
magnetic map of 2013.7 can be utilised to monitor the polarity 
reversal of the star. 

\subsection{The relation between magnetic polarity flips and chromospheric activity
cycle}
\begin{figure}
\centering
\includegraphics[width=0.52\textwidth]{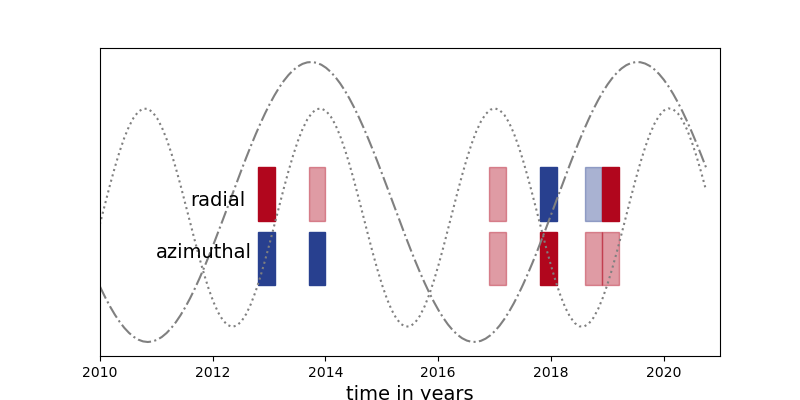}
\caption{Relation between the 5.7 year (dotted 
and dashed grey curve) and 3.1 year (dotted grey
curve) chromospheric activity cycles, 
and the polarity reversals in the radial (top row of rectangles) and 
azimuthal magnetic field (bottom row of rectangles). The red rectangles represent dominant
positive polarity magnetic field and the blue rectangles represent dominant negative
polarity magnetic field. The transparent rectangles mark epochs when additional opposite
polarity magnetic regions are also detected in addition to the dominant polarity.}
\label{corrpolarity}
\end{figure}

The ZDI magnetic maps of $\kappa$ Ceti exhibit polarity reversals
of the radial and the azimuthal magnetic field over a six year time-span.
As shown in Fig.~\ref{corrpolarity},
during this time period the star would undergo at least two 3.1 year (shorter) 
chromospheric cycles and one 5.7 year (longer) cycle. Depending on which 
cycle period one considers only two/one ZDI maps were observed
at an activity maximum (3.1 year period/5.7 year period), none were 
observed at activity minimum, and the rest were observed in between activity
minimum and maximum.

Figure~\ref{corrpolarity} shows that irrespective 
of which cycle period one considers the star's
magnetic evolution deviates from the solar case.
In the Sun a complex magnetic field 
appears on the surface during solar cycle maximum 
and the magnetic field changes polarity from one 
cycle minimum to the next \citep{hathaway10,derosa12}. 
However in the case of $\kappa$ Ceti, the polarity 
flips do not occur from one minimum to the next. Although, 
the complexity of the two maps (appearance of bipolar magnetic
regions) observed during activity maxima 
increases compared to the other epochs.
Reconstructing its magnetic field geometry during cycle minima will
provide stronger constraints on $\kappa$ Ceti's magnetic cycle 
evolution.

Future high cadence observations of this star could help us 
understand the true relationship between the magnetic 
polarity flips and the chromospheric
cycles.
As an example, in the 
case of the planet hosting star $\tau$ Boo, 
initial multi-epoch observations at 
yearly intervals suggested a $\sim$2 year 
magnetic cycle \citep{donati08,fares09,fares13}, at an 1:3 ratio with
the star's chromospheric cycle \citep{mengel16,mittag17}. Additional 
spectropolarimetric observations of the star at monthly intervals 
by \citet{jeffers17} confirmed that its large-scale magnetic field 
reverses polarity more frequently, leading to a magnetic cycle of
240 days. The authors also detected an increase in complexity of 
$\tau$ Boo's magnetic field as it approaches chromospheric activity maximum.

With a rotation period of 9.2 days, 
$\kappa$ Ceti is a moderately rotating young Sun for its age, as shown by stellar rotational
evolution models \citep{gallet13,johnstone15b}. Until now the 
magnetic field evolution 
of only one other moderately rotating young Sun, 
$\epsilon$ Eridani, has been investigated using both chromospheric
activity and ZDI magnetic maps over multiple epochs. Long-term 
monitoring of $\epsilon$ Eridani shows the presence of two
chromospheric cycle periods \citep{metcalfe13}. However, recent investigations on the 
star's chromospheric measurements suggest a complex activity with 
the disappearance of the star's longer cycle period 
\citep{coffaro20,petit21}, indicating strong similarities with $\kappa$ Ceti's chromospheric 
activity evolution. 
The large-scale magnetic field evolution of $\epsilon$
Eridani is more complex than that of $\kappa$ Ceti, although there is
indication of a possible polarity switch in the dipole field \citep{jeffers14,jeffers17,petit21}.
There is a strong possibility that both $\kappa$ Ceti and $\epsilon$ Eridani reflect a magnetic 
evolution that is unique to moderately rotating young Suns.
High-cadence spectroscopic and spectropolarimetric 
observations could help us correctly constrain the evolution of 
$\kappa$ Ceti's large-scale field, which will be addressed in a future work \footnote{A multi-epoch 
high cadence observation programme was recently granted as part of the EU H2020 OPTICON
transnational access programme \url{https://www.astro-opticon.org/h2020/tna/}}.
\section{Summary and conclusions}\label{sec5}
The multi-epoch spectroscopic and spectropolarimetric observations
carried out in this work enabled us to provide an in-depth analysis 
of the complex magnetic variability of the young Sun $\kappa$ Ceti.
The key conclusions are discussed below.
\begin{itemize}
\item The chromospheric activity measurements 
of $\kappa$ Ceti over $\sim$50 years indicate the presence of
two activity cycles with a 1:2 ratio similar to the Schwabe and the Hale cycle. However, unlike
the solar cycles the 3.1 and 5.7 year cycle periods show a complex temporal evolution. 
While the longer cycle dominates for a good portion of our time series, the 
shorter cycle period is appears in the very early and later part of the data set.
Similar complex chromospheric variability is also reported for another moderately rotating young 
star $\epsilon$ Eridani. Our results suggest that such complex evolution of magnetic activity 
could be synonymous with moderately active young Suns, an evolutionary path that our own Sun 
could have taken. 
\item Our ZDI reconstructions show that the radial and azimuthal directions 
of the field undergo polarity reversals,
indicating a potential Solar-like dynamo cycle, with a 
cycle period of $\sim$10 years. 
However, the exact length of the cycle period could vary by a year or two, 
as our current spectropolarimetric observations do not have
a high cadence or a long time-baseline.
\item Magnetic polarity reversals in correlation with chromospheric activity 
is one of the best known ways of constraining the dynamo cycle a Sun-like star.
For a solar-type dynamo the magnetic polarity reversals should coincide with the 
activity cycle minima, followed by the appearance of a complex field during 
activity maxima. Although $\kappa$ Ceti's magnetic field appears to be more bipolar
during activity maxima, the polarity reversals are out of sync with the cycle minima.
Our multi-epoch monitoring suggests that the star's magnetic cycle deviates from the 
Sun-like polarity reversals. 
Further high cadence spectroscopic
and spectropolarimetric observations of cool stars will be crucial in determining if the 
magnetic field in moderately
rotating young Solar analogues indeed evolve differently from the current Sun.
\end{itemize} 
\begin{acknowledgements}
We thank Antoaneta Avramova for valuable discussions on period analysis. 
This work was funded by the Austrian Science Fund's (FWF)
Lise Meitner project M 2829-N, and the FWF NFN project S11601-N16 and S11604-N16.
AA acknowledges the support of the Bulgarian National Science Fund under contract DN 18/13-12.12.2017.
SVJ acknowledges the support of the DFG priority program SPP 1992 “Exploring the Diversity of Extrasolar Planets 
(JE 701/5-1).
AAV acknowledges funding from the European Research Council (ERC) under the European Union's Horizon 
2020 research and innovation programme (grant agreement No 817540, ASTROFLOW). 

The {HK$\_$Project$\_$v1995$\_$NSO} data used in this work derive from the Mount Wilson Observatory 
HK Project, which was supported by both public and private funds through the 
Carnegie Observatories, the Mount Wilson Institute, and the 
Harvard-Smithsonian Center for Astrophysics starting in 1966 and continuing for over 36 years.  
These data are the result of the dedicated work of O. Wilson, A. Vaughan, G. Preston, D. Duncan, S. Baliunas, and many others. 
\end{acknowledgements}
\bibliographystyle{aa}
\bibliography{ref}
\begin{appendix}
\begin{figure*}
\section{}
\includegraphics[width=1\textwidth]{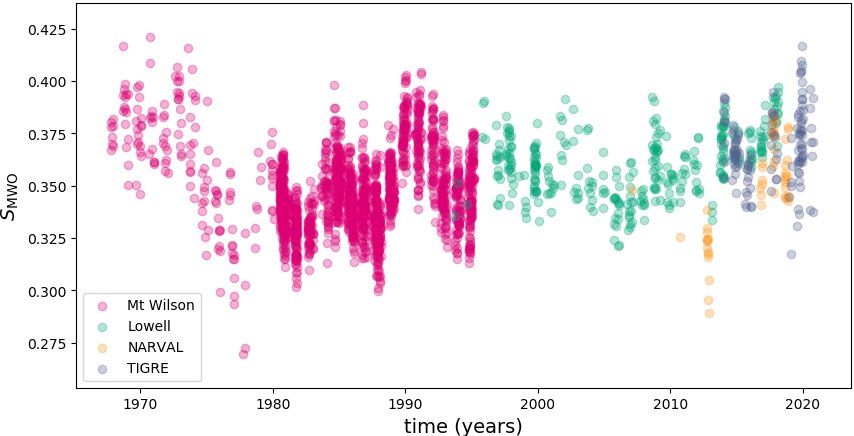}
\caption{Individual $S_\mathrm{MWO}$ measurements of 
$\kappa$ Ceti between 1966-2020. The different colours mark the data sources.}
\label{appendixS}
\end{figure*}

\begin{figure}
\includegraphics[width=.5\textwidth]{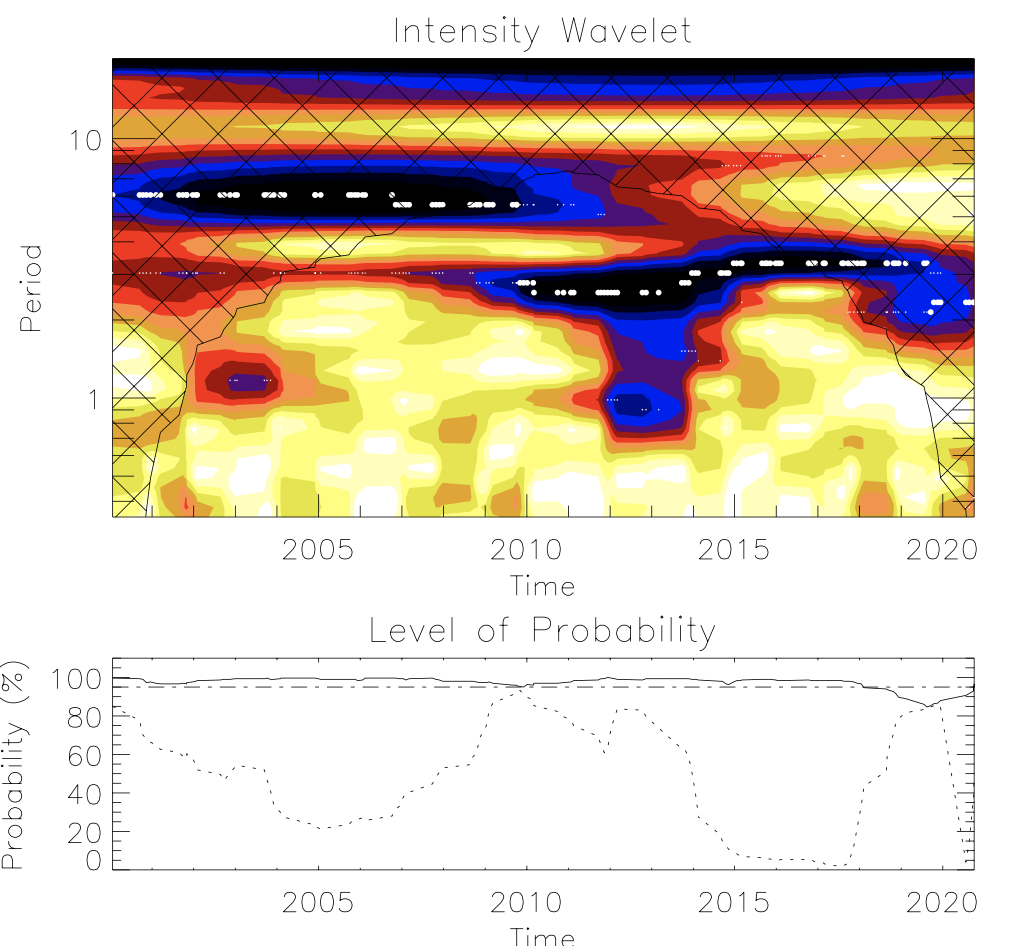}
\caption{Same as Fig.~\ref{waveletfig}, but only for the new data set between 2000-2020. The
solid line represents the probability level for the 3.1 year period and the dotted line 
shows the probability level of the 5.7 year period.}
\label{newave}
\end{figure}
%
\longtab[1]{
\section{Journal of observations.}
\begin{longtable}{cccccc}
\caption{\label{journal} Journal of spectropolarimetric 
observations taken at HARPSpol (2013.7) and NARVAL/TBL 
(2012.8, 2016.9, 2017.8, 2018.6, 2018.9). 
}\\
\hline\hline
Epoch& date & Julian date &$\sigma_\mathrm{LSD}$&rotational\\
&&(2450 000+)&10$^{-5} I_\mathrm{c}$&cycle\\
\endfirsthead 
\caption{continued.}\\ 
\hline\hline 
Epoch& date & Julian date & $\sigma_\mathrm{LSD}$&rotational\\
&&(2450 000+)&10$^{-5} I_\mathrm{c}$&cycle\\
\hline 
\endhead 
\hline 
\endfoot 
\hline
&01 October 2012&6202.52180 &4.8&0.000\\
&02 October 2012&6203.51912 &3.8&0.108\\
&03 October 2012&6204.52650 &6.8&0.218\\
&04 October 2012&6205.53850 &4.2&0.328\\
&05 October 2012&6206.52522 &4.9&0.435\\
&12 October 2012&6213.55626 &7.0&1.199\\
&13 October 2012&6214.48132 &4.5&1.300\\
2012.8&23 October 2012&6224.54605 &6.2&2.394\\
&28 October 2012&6229.55813 &4.9&2.939\\
&31 October 2012&6232.50488 &5.1&3.259\\
&06 November 2012&6238.57330&4.1&3.919\\
&12 November 2012&6244.49623&5.8&4.562\\
&14 November 2012&6246.53620&4.3&4.784\\
&22 November 2012&6254.47162&6.6&5.647\\
\hline
&09 September 2013&6545.7018&9.5&37.302\\
&09 September 2013&6545.9281&8.6&37.327\\
&10 September 2013&6546.7004&20.1&37.435\\
&10 September 2013&6546.9248&6.5&37.411\\
&11 September 2013&6547.7017&11.5&37.544\\
2013.7&11 September 2013&6547.9248&10.8&37.520\\
&12 September 2013&6548.7013&9.1&37.628\\
&12 September 2013&6548.9226&11.7&37.652\\
&15 September 2013&6551.7258&15.9&37.957\\
&15 September 2013&6551.8868&12.6&37.974\\
&15 September 2013&6551.8980&17.5&37.976\\
\hline
&01 November 2016&7694.51937&5.0&162.174\\
&02 November 2016&7695.41739&4.6&162.271\\
&01 December 2016&7724.41020&5.7&165.423\\
2016.9&02 December 2016&7725.33896&5.1&165.524\\
&03 December 2016&7726.31571&4.9&165.630\\
&05 December 2016&7728.32758&4.7&165.848\\
\hline
&26 September 2017&8023.60956 &7.8&197.944\\
&28 September 2017&8025.55739 &7.6&198.156\\
&02 October 2017 &8029.57291  &6.3&198.593\\
&03 October 2017 &8030.61015  &8.0&198.705\\
&04 October 2017 &8031.57700  &10.3&198.810\\
&05 October 2017 &8032.59530  &5.3&198.921\\
2017.8&06 October 2017 &8033.61134  &5.2&199.031\\
&07 October 2017 &8034.57082  &4.9&199.136\\
&09 October 2017 &8036.58561  &5.8&199.355\\
&10 October 2017 &8037.61879  &5.3&199.467\\
&11 October 2017 &8038.61716  &5.6&199.576\\
&12 October 2017 &8039.53068  &6.8&199.675\\
&13 October 2017 &8040.53481  &5.6&199.784\\
\hline
&20 August 2018& 8351.59545 &3.9&233.597\\
&20 August 2018& 8351.61006 &5.2&233.595\\
&21 August 2018& 8352.60425 &4.2&233.705\\
2018.6&22 August 2018& 8353.60739 &3.6&233.815\\
&22 August 2018& 8353.62129 &4.9&233.814\\
&23 August 2018& 8354.54734 &5.0&233.916\\
&31 August 2018& 8362.60629 &2.9&234.792\\
\hline
&24 October 2018& 8416.57808  &5.6&240.658\\
&25 October 2018& 8417.49351  &2.9&240.758\\
&07 November 2018& 8430.53761 &5.1&242.177\\
2018.9&07 November 2018& 8430.55348 &3.3&242.176\\
&13 November 2018& 8436.42431 &5.2&242.815\\
&14 November 2018& 8437.44667 &3.2&242.927\\
&16 November 2018& 8439.55539 &4.2&243.156\\
\hline
\end{longtable}
\tablefoot{
From Left to Right: 
epoch, date of observations, Heliocentric Julian date, the error bar in 
Stokes {\it V}  LSD profile, and the rotational cycles. The rotational cycles
were generated using the parameters listed in Table~\ref{tableproperties}.
}}
\end{appendix}
\end{document}